Risks for Academic Research Projects: An Empirical Study of Perceived Negative Risks and Possible Responses

P. Alison Paprica (corresponding author)

Institute for Health Policy, Management and Evaluation

University of Toronto

55 College St 4th Floor, Toronto, ON M5T 3M6

alison.paprica@utoronto.ca

416-802-2651

Abstract

Academic research projects receive hundreds of billions of dollars of government investment each year. They complement business research projects by focusing on the generation of new foundational knowledge and addressing societal challenges. Despite the importance of academic research, the management of it is often undisciplined and ad hoc. It has been postulated that the inherent uncertainty and complexity of academic research projects make them challenging to manage. However, based on this study's analysis of input and voting from more than 500 academic research team members in facilitated risk management sessions, the most important perceived negative risks were general, as opposed to being research specific. Overall, participants' top risks related to funding, team instability, unreliable partners, study participant recruitment, and data access. Many of these risks would require system- or organization-level responses that are beyond the scope of individual academic research teams.

1. Introduction

In order to generate new knowledge and address societal challenges, countries around the world are making major investments in academic research – defined here as research performed at universities, research hospitals and other public research institutes. It has been postulated that that management of academic research projects is particularly challenging given the inherent uncertainty of research and the need for risks to be taken if innovation is to be achieved (Ernø-kjølhede, 2000) [1]. However, the research literature on academic research project leadership and management is sparse, particularly on the topic of risk management (vom Brocke and Lippe, 2015; Moore and Shangraw, 2015) [2,3]. The objective of this study was to learn which specific negative risks for academic research projects are perceived to be important by academic research team member, and what team members see as possible risk responses.

2. Method

This study involved retrospective analyses of data from a convenience sample of over 500 participants of in-person and online facilitated sessions focused on risk management for academic research (Table 1). The first 1.5-hour workshop involved a large group (estimated 200+ research support staff) at a concurrent session presentation entitled "Risk Management for Research" at the 2015 Canadian Association of Research Administrators (CARA) Conference in Toronto, Canada.

Fourteen (14) additional facilitated risk management sessions, with a collective total of 314 participants, were conducted as part of research project management courses and workshops. Most course and workshop participants were Canadian but the study also includes input from academic research team members from the UK, Europe and Africa.

The roles, number, and primary discipline of participants varied across the risk management sessions (Box 1, Table 1). Each of the 15 facilitated sessions was one to 1.5 hours in length and followed the same general steps which were: brief seminar, individual brainstorming of negative risks, group voting and discussion to identify the top negative risks, and the development of possible risk responses for key risks (Box 2).

---

Box 1. Roles of Academic Research Team Member Participants

- Researchers – Principal Investigators (PIs) and other research team members with academic appointments

- Research staff – staff scientists, research associates, statisticians, technicians, and other staff who contribute research and scientific expertise through paid staff positions

- Project managers – staff who have the responsibility for project management, noting that the title may be executive director, project director, project manager, project coordinator, or some other title depending on the institution and size and scope of the research

- Fellows – researchers who hold time-limited fellowship positions and have completed their doctoral work (postdocs) or achieved other discipline-specific degrees (e.g., medical doctor)

- Graduate Students – individuals performing thesis research to fulfill PhD or Master's degree requirements (note: undergraduate students may also be part of a research teams but none were included in this study)

- Support staff – research administrators, grant officers, technology transfer experts, legal counsel, and other staff who directly support the planning, administration, and implementation of academic research

---

> Box 2: Process for Facilitated Risk Management Sessions
>
> 1) Ten to 15-minute seminar on the theory of risk management for academic research projects
>
> 2) Generation of a list of negative risks
>
>    a) Individual brainstorming of negative risks
>
>    b) Depending on the size of group and length of session, work in pairs to identify negative risks that may be important based on simultaneous consideration of risk likelihood and impact
>
> 3) Group discussion to generate a session risk registry that includes some, but not all, of the individually brainstormed risks
>
> 4) Individual voting to identify the top risks within the session risk registry, taking both likelihood and impact into account
>
> 5) Group discussion to compare the session's top risks with the top risks from other sessions (with the exception of the first session at the CARA Conference)
>
> 6) Participants working together (as a large group or in small breakout groups depending on the session format and duration) to develop possible risk responses (e.g., avoid, mitigate, transfer, accept) for two or three of the session's top risks.

Most data were captured using live online polling (Poll Everywhere) without any identifying information about participants. In other cases, brainstormed risks were recorded and displayed using projected computer screens, white boards, flip charts, or sticky notes and voting was accomplished through the placement of individual stickers on flip charts or a show of hands.

Data were prepared for analysis by assigning risk categories and subcategory labels to individual risks (e.g., Funding – budget cut) to facilitate the identification of common risks and themes across sessions. The top three to five negative risks with the most votes were identified for each session (Table 1). In risk management sessions with more than 20 participants, there was often a natural clustering of three to five risks with a large number of votes, followed by a large number of risks with significantly fewer votes, but this was not always observed. In cases where there was not an obvious cluster of top risks with many votes, the three risks with the most votes were included in the analysis, or the top four risks, if there was a tie for third.

3) Theory

The total higher education research and development (HERD) performed by universities, research hospitals, colleges, and other public and not-for-profit research institutes in Canada in fiscal year 2018/19 was more than $15 billion CAD [4]. Appreciating that it can be difficult to understand such large currency values; for reference, the 2018/19 Canadian HERD is almost twenty times the annual value of Canadian logging exports ($777 million CAD in 2019) [5] and more than five times the total annual revenue of all performing arts businesses in Canada ($2.4 billion CAD in 2018) [6]. Notwithstanding that Canadian HERD represents a significant investment, it is less than ten per cent of the United States' $83.7 billion USD HERD in 2019 [7], and a minor contributor to the estimated total OECD HERD of $248 billion USD-equivalent based on 2018 data or the most recent year's HERD data available for each OECD country [8].

Academic research is primarily funded by governments and plays an extremely important role in innovation systems by ensuring the provision of new knowledge from basic and applied research that private firms are unlikely to conduct because of the non-appropriable, public good, intangible character of knowledge and the risky nature of research (OECD, 2012) [9]. The specific objectives of the funders of academic research vary, but generally include the generation of new foundational knowledge and/or research findings that will directly or indirectly lead to social, health, environmental and economic benefits (OECD, 2016) [10]. A recent trend in publicly funded academic research is the mobilization of large interdisciplinary teams to address societal challenges as exemplified by the European Horizon 2020 Program [11], the UK Global

Challenges Program [12], and the Canadian New Frontiers in Research Fund – Transformation competition [13].

Though the size and importance of HERD investment around the world is large, managerial practice in the academic sector is undisciplined, ad hoc, with some research leaders being openly "anti-management" (vom Brocke and Lippe, 2015; Moore and Shangraw, 2015) [2,3]. Within the limited research literature focused on the leadership and management of academic research projects, there is general agreement that the nature of research projects necessitates different approaches to project management than are used for traditional projects in other sectors. Many authors cite or paraphrase Ernø-kjølhede's statement from over twenty years ago: "The management of a research project is full of uncertainty and complexity. Research has substantial elements of creativity and innovation and predicting the outcome of research in full is therefore very difficult." (Ernø-kjølhede, 2000) [1] Recent peer-reviewed articles on focus mostly on frameworks and tailored approaches that aim to modify conventional approaches so that they work for academic research projects (vom Brocke and Lippe, 2015; Kuchta et al, 2017; Powers and Kerr, 2009; Cassanelli et al., 2017; Huljenic et al, 2005) [2,14,15,16,17].

Within project management, risks are understood to be uncertain events or conditions that, if they occur, would have a positive or negative effect on one or more project objectives (Wang et al, 2010; Project Management Institute, 2017) [18,19]. In practice, most project risk management focuses on identifying important negative risks, analyzing negative risks to assess likelihood and impact, developing negative risk responses (e.g., avoid, mitigate, transfer, accept), and monitoring and controlling negative risks during project implementation (Project Management Institute, 2017) [19]. Risk management has been identified as one of most challenging aspects for academic research projects because the inherent uncertainty of research projects hinders risk identification, risk response planning and risk monitoring (Moore and Shangraw, 2015; Kuchta et al. 2017) [3,14]. However, the literature on risk management for academic research projects is extremely sparse and offers little information about which specific negative risks warrant responses because they pose threats to project success. Two recent review articles with a focus on academic research project management (vom Brocke and Lippe, 2015; Philbin, 2017]) [2,20] identified three relevant risk-focused publications but did not report specific risks that academic research projects may need to manage. One relevant publication cited in the review articles was a framework that describes how different risk management frameworks could be brought together to manage risks at many levels at the National Research Council of Canada (Leung et al, 2008) [21]. The review articles also identified two empirical studies with a focus on risk: one qualitative study that included risk-focused interviews with an unspecified number of project managers from an enterprise in Malaysia (Ibrahim et al, 2013) [22], and one qualitative study that focused on the risks and rewards perceived by 30 people involved in Australian Cooperative Research Centres (CRCs) (Garrett-Jones et al., 2005) [23]. Other publications cited in the review articles describe practice changes that could help manage any uncertain project (e.g., planning in phases).

Investigation of literature that cited, or was cited by, the references noted above yielded two additional publications that directly describe specific risks for academic research projects: Moore and Shangraw's 2011 analysis of 18 partial responses and 12 complete responses to a 28-page survey sent to 58 project managers of large government funded research projects [3], and Bodea and Dascalu's 2009 article focused on computer-aided methods for risk analysis which includes a list of 19 risks for research projects developed by the authors without explicit reference to the evidence used to generate their list of risks [24]. Given the large investment in academic research, and the potential importance of academic research for societal problems, additional research is needed to understand, and develop responses for, risks for academic research projects.

4) Results and Discussion

4.1 Results

Table 1: Top Perceived Negative Risks in 15 Facilitated Risk Management Sessions

| No. | Year | Partici-pants | Context in Which Facilitated Session was Provided (in-person unless noted) | Top Negative Risks (in order of perceived importance based on participant voting) |
|---|---|---|---|---|
| 1 | 2015 | >200 | 1.5-hour workshop for research administrators and other support staff | 1.Contractual – non-compliance (financial fraud or scope not delivered)<br>2.(tie) Funding – budget cut<br>2.(tie)Team – staff member leaves |
| 2 | 2015 | 6 | Part of a 25-hour health sciences graduate student course | 1.Data – delayed access<br>2.(tie) Sample – delays with recruitment<br>2.(tie) Schedule – delayed approval to start |
| 3 | 2017 | 14 | Part of a 6-hour workshop for natural sciences and social sciences graduate students and fellows | 1.Team – team member leaves<br>2.Schedule – overly optimistic<br>3.Funding – funder withdraws |
| 4 | 2017 | 31 | Part of a 6-hour workshop for health sciences fellows and PhD students | 1.Partner – unresponsive<br>2.Partner – doesn't make needed contributions<br>3.Team – team member leaves<br>4.Funding – grant proposal not funded/renewed<br>5.Sample - underpowered/insufficient |
| 5 | 2017 | 8 | Part of a 25-hour health sciences graduate student course | 1.(tie) Data – delayed access<br>1.(tie) Sample - underpowered/insufficient<br>1.(tie) Study– unable to retain participants<br>1.(tie) Schedule – delayed approval to start |
| 6 | 2018 | 42 | Part of a 6-hour workshop for health sciences fellows and PhD students | 1.Data – insufficient quality<br>2.Partner – lack of buy-in<br>3.Data – delayed access<br>4. Sample - underpowered/insufficient<br>5.Partner – doesn't make needed contributions |
| 7 | 2018 | 25 | Part of a 36-hour continuing education course for researchers, research staff, PMs, fellows, graduate students, and support staff from a range of disciplines | 1.Partner – unresponsive<br>2.Partner – doesn't make needed contributions<br>3.Funding – grant proposal not funded/renewed<br>4.(tie) Team – team member leaves<br>4.(tie) Sample - underpowered/insufficient |

| No. | Year | Partici-pants | Context in Which Facilitated Session was Provided (in-person unless noted) | Top Negative Risks (in order of perceived importance based on participant voting) |
|---|---|---|---|---|
| 8 | 2018 | 22 | Part of a 3-hour workshop at an international academic conference for researchers, research staff, fellows, and graduate students from a range of disciplines | 1. Data – delayed access<br>2. Partners – doesn't make needed contributions<br>3.(tie) Data – insufficient quality<br>3.(tie) Context –urgent issues crowd out research |
| 9 | 2018 | 10 | Part of a 6-hour workshop for natural sciences researchers, research staff, fellows, and graduate students | 1. Team – supervisor or PI leaves<br>2. Team – interpersonal conflict<br>3. Research – doesn't produce conclusive results |
| 10 | 2018 | 46 | Part of a 15-hour workshop for natural sciences researchers, research staff, PMs, and support staff | 1. Funding – budget cut<br>2. Funding – delayed start<br>3.(tie) Team – lacks essential skills<br>3.(tie) External – climate/environmental risks |
| 12 | 2019 | 19 | Part of a 6-hour workshop for researchers, research staff, PMs, fellows, graduate students, and support staff from a range of disciplines | 1. Sample - underpowered/insufficient<br>2.(tie) Funding – budget cut<br>2.(tie) Data – delayed access<br>2.(tie) Schedule – delayed approval to start |
| 12 | 2019 | 41 | Part of a 6-hour workshop for health sciences fellows and PhD students | 1. Partner – lack of buy-in<br>2. Data – delayed access<br>3.(tie) Sample - underpowered/insufficient<br>3.(tie) External – policy/political uncertainty |
| 13 | 2019 | 19 | Part of a 6-hour workshop for researchers, research staff, PMs, fellows, graduate students, and support staff from a range of disciplines | 1.(tie) Team – team member leaves<br>1.(tie) Study – undetected error in analysis<br>1.(tie) Equipment – failure to function |
| 14 | 2020 | 20 | Part of a 15-hour online workshop for natural sciences researchers, research staff, PMs, and support staff | 1. Funding – budget cut<br>2.(tie) Team – staff member leaves<br>2.(tie) Contractual – non-compliance (financial fraud or scope not delivered) |

| No. | Year | Partici-pants | Context in Which Facilitated Session was Provided (in-person unless noted) | Top Negative Risks (in order of perceived importance based on participant voting) |
|---|---|---|---|---|
| 15 | 2021 | 13 | Part of a 6-hour online workshop for researchers, research staff, PMs, fellows, graduate students, and support staff from a range of disciplines | 1.External – another pandemic<br>2.(tie) Team – team member leaves<br>2.(tie) Funding – funding runs out |

Risks related to funding and the project team were the most common negative risks to be identified as important based on participant votes. Eight groups voted one or more risks related to funding into their short list of top risks. Usually, these groups focused on the risk of budget cuts, or the risk that funders would withdraw, but two groups identified the risk that grant funding would not be approved or renewed, and one group identified delays in funding as a negative risk that warranted a response even if the funds were eventually received. These results were consistent with the findings of Moore and Shangra who reported that only one project manager (out of five respondents to that question) reported their large research project being completed on budget, and that 57 per cent of respondents indicated that they experienced staff turnover (Moore and Shangra, 2011) [3].

When participants of facilitated sessions were given the option of choosing specific risks to develop responses for, funding risks were the most popular choice across the 15 facilitated sessions. Suggested responses for funding risks generally included: (i) (mitigate likelihood) build and maintain strong personal relationships with the funder, (ii) (mitigate impact) invest time and resources in identifying additional alternative funders, (iii) (mitigate likelihood) incorporate and highlight milestones and deliverables that clearly and obviously align with the funder's preferences and needs, and/or (iv) (mitigate impact) proactively identify the activities and deliverables that will be delayed, cut, or partially reduced if negative risks related to funding are realized.

Eight of the 15 sessions voted the risk that a team member would leave or be unavailable as one of their top risks. In some cases, participants focused on the risk that a team would become short-staffed if a staff person were hired away, in other cases the concern was that the Principal Investigator or a research staff person would leave the project (temporarily or permanently) or become ill or die, others referred to general issues with turnover. After funding risks, the risk of a team member leaving or being unavailable was the second most frequently selected risk for risk response development during facilitated sessions. Proposed responses to this risk generally included: (i) (mitigate impact) encourage or require people to put important information in documents that others can access (ii) (mitigate impact) require team members with highly specialized skills to train or mentor at least one other person on the team, (iii) (mitigate impact) in cases where an individual has a planned departure date, reserve their last two weeks for knowledge transfer activities and/or (iv) (mitigate likelihood) offer a flexible work environment that is interesting, rewarding, and respectful of all team members so that people are less likely to look for work elsewhere.

Other negative risks that were identified as important in multiple sessions included, seven groups identifying risks associated with sample size (predominantly the risk that studies would not be able to recruit or retain a sufficient number of participants) and six groups identifying risks associated with data (predominantly the risks that access would be delayed or that data quality would be insufficient). Five groups identified the risk that a partner or stakeholder would lose interest, become unresponsive or not deliver their planned contributions to the project. Several groups opted to develop potential responses to partner-related risks, identifying responses that were similar to the responses to funding risks in that they focused on building and maintaining relationships, and paying careful attention to fulfilling partner needs.

Risk of contractual non-compliance (e.g., teams not producing deliverables specified in the contract and/or committing financial fraud) was identified as a top risk in just two session; but it noteworthy because it was perceived to be one of the most important risks for academic research by almost all of the 200+ research administrators and support staff at a large group facilitated session.

Study participants did identify some risks that might be considered inherently associated with academic research, but these risks did not receive sufficient votes to be included among the top risks in Table 1. These "inherent" risks included: another group publishes findings before the research was completed (getting "scooped"); unintentional harm to research study participants; research that does not yield, meaningful, reproduceable, or publishable results; Research Ethics Board/Institutional Review Board approval is withheld or withdrawn due to safety concerns; and the risk that the technology needed to perform the research does not exist.

4.2 Discussion

Overall, the risks that were perceived to be the most important risks for academic research projects were general risks as opposed to risks directly associated with inherent uncertainty or complexity of academic research. Across multiple facilitated sessions, there were commonalities with many participants identifying unstable funding, high staff turnover, and unreliable partners as one of their top risks. Notably, these are significant risks that could consume the attention and affect the work of anyone working in any sector, not just academic research.

The study also identified risks that may be more closely associated with research projects than non-research projects such as risks related to participant recruitment, sample size, data access and data quality. However, it should be noted that these risks are not unique to academic research and could also affect business research and a range of non-research activities such as corporate quality improvement initiatives and public consultations conducted by governments and government agencies.

The top risks identified by participants of this study are not aligned with statements in the limited literature which suggests that the inherent uncertainty and complexity of research are major drivers of risks for academic research. Study participants did identify some risks that might be considered inherently associated with academic research, including the risk of being "scooped" and risk that the research does not yield meaningful, reproduceable or publishable results. Arguably, the risk of contractual non-compliance could be seen as being directly related to the uncertainty of research in that it is a challenge for academic research contracts to accurately forecast work. However, most of the research-related risks that were identified by participants of this study would also be risks for simple, straightforward research projects, not just complex and uncertain academic research projects.

It is possible that the discrepancy between the risks that the literature suggests will be important and what participants perceived to be important occurred because participants are consumed with foundational risks that pose serious threats to their projects and are not yet aware of the risks uniquely associated with academic research. It is also possible that the discrepancy is a by-product of the facilitated session process because risks that were uniquely associated with the complexity or uncertainty of one participant's research project would be less likely to be supported by other and make it onto the group risk registry or short list of top risks. These possible explanations do not mean that set of negative risks perceived by participants is invalid. Rather, the consistency with which the same risks were identified by diverse participants across multiple facilitated risk sessions suggests that many academic research team members perceive them to be important risks that warrant risk responses.

Workshop participants did identify some individual- or team-level responses that could mitigate risks related to funding, team instability and partners. Nevertheless, objectively, it seems that the most important negative risks perceived by participants will require system- and organization-level management responses and remedies. For example, it is striking how many groups independently identified unstable funding as a top risk. Though a research team might be able to mobilize support to decrease the likelihood of budget cuts for their individual project, in the absence of research funding reform, individual project funding stability may come at the direct expense of other projects that experience decreases in funding as a direct result. Similarly, there are limits to what a research team or principal investigator can do to mitigate the risk that a staff member will leave the team if the reason for the staff member's departure is that grant funding ended without renewal and there are no funds to pay their salary. Additionally, the negative risks that participants perceived related to unreliable partners are noteworthy in the current context of the trend toward large-scale research grants that require partnerships with industry, government, and other knowledge users. For understandable reasons, some participants perceived such partnerships to create new risks that they do not have the skill set to manage. Changes to research funding strategies, or additional partnership supports, may be required to address these partner-related risks.

4.3 Limitations

This study has limitations. Foremost, it is based on retrospective analysis of a convenience sample of people who self-selected to learn more about project management for research, and the findings may not reflect the views of people who are less interested in research project management training. Secondly, the responses of participants may not be informed or accurate. While it is possible that some of the 200+ participants of the risk session at the 2015 CARA Conference had deep knowledge and expertise related academic project risks, many of the participants of the other 14 workshops and courses were researchers, staff, fellows, and graduate students who are in the early stages of their careers. As such, the findings may not accurately reflect the views and knowledge of more experienced research team members and academic leaders. Most of the participants were Canadians, and the findings may not reflect the views of academic research team members in other countries. Finally, it is not possible to assess the relevance of individual characteristics (e.g., role on the research team, educational background, years of experience, size or nature of research project) because individual-level data were not collected. Additional, prospective individual-level data from research studies with purposive sampling will be required to understand how individual characteristics contribute to risk perception and risk response planning for academic research projects.

5. Conclusions

An analysis of the input and votes of over 500 participants in 15 facilitated risks management sessions found that negative risks related to funding, team instability, unreliable partners, study participant recruitment, and data access were perceived to be the most important negative risks for academic research projects. Overall, most of the negative risks that were perceived to be important were general, as opposed to directly associated with the inherent uncertainty or complexity of academic research. Most of the negative risks that were perceived to be important cannot be fully managed by research teams and would require system- and organization-level responses.

Funding

This research did not receive any specific grant from funding agencies in the public, commercial, or not-for-profit sectors.

References

[1] E. Ernø-kjølhede, Project Management Theory and the Management of Research Projects, Working Papers 3/2000, (2000) Copenhagen Business School, Department of Management, Politics & Philosophy. Google Scholar

[2] J. vom Brocke, S. Lippe, Managing collaborative research projects: A synthesis of project management literature and directives for future research, In Int. J. Proj. Manag. 33- 5 (2015) 1022-1039, https://doi.org/10.1016/j.ijproman.2015.02.001.

[3] S. Moore and R.F. Shangraw Jr, Managing Risk and Uncertainty in Large-Scale University Research Projects, Research Management Review, 18(2) (2011) 59-78. Google Scholar

[4] Statistics Canada, Provincial estimates of research and development expenditures in the higher education sector, by funding sector and type of science https://www150.statcan.gc.ca/t1/tbl1/en/tv.action?pid=2710002501 (accessed 14 Mar 2021)

[5] Government of Canada, Summary - Canadian Industry Statistics – Logging https://www.ic.gc.ca/app/scr/app/cis/summary-sommaire/1133 (accessed 14 Mar 2021)

[6] Statistics Canada, Performing arts, 2018 https://www150.statcan.gc.ca/n1/daily-quotidien/200227/dq200227d-eng.htm (accessed 14 Mar 2021)

[7] National Science Foundation, https://ncses.nsf.gov/pubs/nsf21313/ (accessed 14 Mar 2021)

[8] OECD Main Science and Technology Indicators, Volume 2020 Issue 1 (2020) pg. 52 https://doi.org/10.1787/e3c3bda6-en

[9] OECD Science, Technology and Industry Outlook 2012 – Chapter 6: PUBLIC RESEARCH POLICY (2012) 177-180 https://dx.doi.org/10.1787/sti_outlook-2012-en

[10] Public research missions and orientation, in OECD Science, Technology and Innovation Outlook 2016 (2016) https://doi.org/10.1787/sti_in_outlook-2016-35-en

[11] Horizon 2020 https://ec.europa.eu/programmes/horizon2020/en (accessed 14 Mar 2021)

[12] Global Challenges Research Fund https://www.ukri.org/our-work/collaborating-internationally/global-challenges-research-fund/ (accessed 14 Mar 2021)

[13] New Frontiers in Research Fund, Transformation Competition https://www.sshrc-crsh.gc.ca/funding-financement/nfrf-fnfr/transformation/2020/competition-concours-eng.aspx (accessed 14 Mar 2021)

[14] D. Kuchta, B. Gładysz, D. Skowron, J. Betta, R & D projects in the science sector. R&D Management. Jan;47(1) (2017) 88-110. https://doi.org/10.1111/radm.12158

[15] L.C. Powers and G. Kerr, Gillian, Project Management and Success in Academic Research, REALWORLD SYSTEMS - RESEARCH SERIES. 2009:2 (2009) http://dx.doi.org/10.2139/ssrn.1408032

[16] A.N. Cassanelli, G. Fernandez-Sanchez, M.C. Guiridlian, Principal researcher and project manager: who should drive R&D projects?, R&D Management. 47.2 (2017) 277-287. https://doi.org/10.1111/radm.12213

[17] D. Huljenic, S. Desic, M. Matijasevic, Project management in research projects, Proceedings of the 8th International Conference on Telecommunications, 2005. ConTEL Vol. 2. IEEE (2005) Google Scholar

[18] J. Wang, W. Lin, Y.H. Huang YH, A performance-oriented risk management framework for innovative R&D projects. Technovation. Nov 1;30(11-12) (2010) 601-11 https://doi.org/10.1016/j.technovation.2010.07.003

[19] A guide to the Project Management Body of Knowledge (PMBOK guide) 6th ed, Project Management Institute, 2017

[20] S.P. Philbin, Investigating the Application of Project Management Principles to Research Projects–An Exploratory Study. Proceedings of the 38th American Society for Engineering Management (ASEM) International Annual Conference. (2017) Google Scholar

[21] F. Leung, F. Isaacs, Risk management in public sector research: approach and lessons learned at a national research organization, R&D Management. 38.5 (2008) 510-519 Google Scholar

[22] J. Ibrahim, S. Wani, S., M.E. Adam, H.O. Abdullahi, A.F. Shamsudin, Risk management in parallel projects: Analysis, best practices and implications to DBrain (gDBrain) research project. Fifth Information and Communication Technology for the Muslim World (ICT4M), IEEE (2013) 1-5 Google Scholar

[23] S. Garrett-Jones, T. Turpin, P. Burns, K. Diment, Common purpose and divided loyalties: the risks and rewards of cross-sector collaboration for academic and government researchers. R&D Management. 35.5 (2005) 535-544 https://doi.org/10.1111/j.1467-9310.2005.00410.x

[24] C.N. Bodea, M.I. Dascalu MI, Modeling Research Project Risks with Fuzzy Maps, Journal of Applied Quantitative Methods. 4(1) (2009) 17-30. Google Scholar